\documentstyle[12pt]{report}
\begin{document}
\noindent \today
\def\singlespacing{\baselineskip=12pt}
\def\doublespacing{\baselineskip=24pt}

\doublespacing

 \begin{center}
\begin{large}

{\bf Self-Consistent Theory of Polymerized Membranes}

\end{large}

\bigskip

\bigskip

Pierre Le Doussal$^*$

\medskip

{\it Institute for Advanced Study, Princeton NJ 08540 USA}

\medskip

and

\medskip

Leo Radzihovsky

\medskip

{\it Lyman Laboratory, Harvard University, Cambridge MA 02138 USA}

\end{center}

\medskip

\begin{center}
{\bf Abstract}
\end{center}

We study $D$-dimensional polymerized membranes embedded
in $d$ dimensions using a self-consistent screening approximation.
It is exact for large $d$ to order $1/d$, for any $d$ to
order $\epsilon=4-D$ and for $d=D$. For flat physical membranes ($D=2,d=3$) it
predicts a roughness exponent $\zeta=0.590$.
For phantom membranes at the crumpling transition the size exponent is
$\nu=0.732$.
It yields identical lower critical dimension  for
the flat phase and crumpling transition $D_{lc}(d)={2 d \over {d+1}}$
($D_{lc}={\sqrt{2}}$ for codimension 1).
For physical membranes with ${\it random}$ quenched curvature $\zeta=0.775$ in
the new $T=0$ flat phase
in good agreement with simulations.

\bigskip


\newpage

\bigskip

There are now several experimental realizations of polymerized or solid-like
membranes, such as protein networks of biological membranes$^{1,2}$,
polymerized
lipid bilayers$^{3}$ and some inorganic surfaces$^{4}$.
Unlike linear polymers, two dimensional sheets of molecules with
fixed connectivity and non zero shear modulus are predicted to exhibit a flat
phase with broken orientational
symmetry.
Out of plane thermal undulations
of solid membranes which induce a non-zero local Gaussian curvature
are strongly suppressed because they
are accompanied by in-plane shear deformations$^{5}$.
As a result, even "phantom" tethered membranes
should be flat at low temperatures$^{5,6}$, and exhibit
a quite remarkable anomalous elasticity, with wavevector
dependent elastic moduli that vanish and a bending
rigidity that diverges at long wavelength$^{7}$.
Excluded volume interactions, present
in physical membranes, further stabilize the flat phase$^{8}$
but are usually assumed to be otherwise irrelevant to describe its
long distance properties.
Motivated by recent experiments on partially polymerized
vesicles$^{3}$, studies of
models with quenched in-plane disorder
have shown that the flat phase is unstable at T=0 to
either local random stresses$^{9}$ or random
spontaneous curvature$^{10}$.

\bigskip

Flat membranes of  internal dimensionality  D and linear size L
are characterized by a roughness exponent ${\zeta}$
such that transverse displacements scale as $L^{\zeta}$.
Nelson and Peliti (NP), using a
simple one loop self-consistent theory$^{5}$ for $D=2$ which {\it assumes}
non-vanishing elastic
constants, found that phonon-mediated
interactions between capillary waves lead to a
renormalized bending rigidity $\kappa_R(q) \sim
q^{-\eta}$ with ${\eta=1}$. Since $\zeta=(4-D-\eta)/2$ they predicted
$\zeta=1/2$
for physical membranes.
An $\epsilon=4-D$  expansion$^{7}$ confirmed that
the flat phase was described
by a non trivial fixed point, but with {\it anomalous} elastic constants
$\lambda(q)
\sim \mu(q) \sim q^{\eta_u}$, $\eta_u>0$,
with ${\eta_u=4-D-2 \eta}$ as a consequence of
rotational invariance. Thus, in general $\zeta=(4-D+\eta_u)/4$
and the NP approximation corresponds to setting
${\eta_u}=0$.

\bigskip

There is presently some
uncertainty on the precise value of the roughness exponent for physical
membranes.
Numerical simulations of tethered
surfaces display a range of values for ${\zeta}$
from 0.5$^{2}$, 0.53$^{11}$, 0.64$^{8,12}$ to 0.70$^{13}$.  On the other hand,
the $O(\epsilon)$ result$^{7}$ suggests a value very close to the NP value
$1/2$,
( 0.52 by naively setting $\epsilon=2$). ${\zeta}$ should soon be measured from
experiments,
either directly from light scattering on diluted
solutions$^{4}$ or indirectly, from the scale
dependence of the elasticity$^{14}$ of lamellar stacks of solid membranes
presently
under experimental study. The buckling transition$^{6}$, if observed,
is controlled by a single exponent related to ${\zeta}$. It thus seems
desirable
to explore further possible theoretical predictions for ${\zeta}$.

\bigskip

In this Letter we introduce a self-consistent approximation which improves on
the Nelson Peliti
theory$^{5}$ by allowing a non trivial renormalization of the elastic moduli.
It is exact in three different limits and compares well with numerical
simulations.
We construct two coupled self-consistent equations for the renormalized bending
rigidity
$\kappa_R(q)$ and elastic moduli $\mu_R(q),\lambda_R(q)$ and solve them in the
long
wavelength limit. $\kappa_R(q)$ is determined by the propagator
for the  $d_c=d-D$ components ${\bf h}$ of the out of plane fluctuations
$G(q)\sim 1/q^{4-D-\eta}$ while the
elastic moduli are determined by the four-point correlation function of ${\bf
h}$ fields.
Physically, our calculation includes the additional effect of relaxation of
in-plane
stresses by out of plane displacements. As a result, curvature fluctuations
soften elastic constants
and screen the phonon-mediated interaction.
A similar Self-Consistent Screening Approximation (SCSA) was introduced
by Bray$^{15}$ to estimate the $\eta$
exponent of the critical $O(n)$ model (here $d_c$ plays
the role of the number of components $n$)
and amounts to a partial resummation of the $1/d_c$ expansion.
By construction, the method is exact for
large codimension $d_c$ to first order in $1/d_c$ and arbitrary $D$.
Solving self-consistently then leads to an improved approximation of
$\eta(d_c,D)$
(and thus $\zeta$)
for the small (physical) values of $d_c$.

\bigskip

The attractive feature of our theory is that it becomes exact in several other
limits.
Firstly, because of the Ward identities associated to rotational invariance
we find that $\eta(d_c,D)$ is exact to first order in $\epsilon=4-D$ for
arbitrary $d_c$
and is thus compatible with all presently known results$^{6,7}$.
Secondly, for $d_c=0$ it gives $\eta=(4-D)/2$ which is the exact result
since clearly $\eta_u=0$ for $d=D$, and$^{7}$ ${\eta_u=4-D-2 \eta}$.
This is at variance with the $O(n)$ model for which the SCSA$^{15}$ is not
exact for $n=0$.
Thus we expect this method to give more accurate results for the
present problem. Two-loop calculations are in progress$^{16}$ to estimate the
deviation. An encouraging indication
is the similarity of our method with the remarkably accurate self-consistent
approximation
of Kawasaki$^{17}$ for the critical dynamics of the binary fluid mixture,
which was shown to be exact to order $\epsilon$, again because of Ward
identities,
and incorrect to order $\epsilon^2$ by a tiny amount.
We also apply this method to the crumpling transition of phantom
membranes, and to flat membranes with quenched disorder. Details can be found
in Ref.16.

\bigskip

In the flat phase, the membrane in-plane
and out-of plane displacements are parametrized respectively
by a D-component phonon field $u_{\alpha}(x)$, $\alpha=1,..D$,
and a $d_c=d-D$ component
out-of plane height fluctuations field ${\bf h}(x)$. A monomer of internal
coordinate $x$ is at position $
{\bf r}(x) = (x_{\alpha} + u_{\alpha}(x) ) {\bf e}_{\alpha} +  {\bf h}(x)
$
where
the ${\bf e}_{\alpha}$ are a set of D orthonormal vectors.
The effective free energy is the sum of a bending energy and an in plane
elastic energy (most relevant terms):
$$
F= \int d^Dx  [ {\kappa \over 2} (\nabla^2{\bf h})^2
 + \mu u_{\alpha \beta}^2 +{\lambda \over 2}
u_{\alpha \alpha}^2 ]
\eqno(1)
$$
where the strain tensor is $ u_{\alpha \beta}=
 {1\over 2} ( {\partial _{\alpha} u_{\beta}} +
{\partial _{\beta} u_{\alpha} } +
{\partial _{\alpha} {\bf h}}\cdot{\partial _{\beta} {\bf h}} )$
To discuss the SCSA in the flat phase it is convenient to first
integrate out the phonons$^{1,5}$, and to work with the
$d_c$-component ${\bf h}$ field. In terms of Fourier components the free energy
takes
the form of a critical theory:
$$
F_{eff}={\kappa \over 2} \int dk  k^4 \mid {\bf h}(k) \mid^2 +
{1 \over {4 d_c} } \int dk_1dk_2dk_3 R_{\alpha \beta, \gamma \delta}(q)~
k_{1 \alpha} k_{2 \beta} k_{3 \gamma} k_{4 \delta}~
{\bf h}(k_1).{\bf h}(k_2 )~
{\bf h}(k_3).{\bf h}(k_4 )
\eqno(2)
$$
with $q=k_1+k_2$ and $k_1+k_2+k_3+k_4=0$ and we use $\int dk$ to denote $\int
d^Dk/(2 \pi)^D$.
The four-point coupling fourth-order tensor $R(q)$ is transverse to $q$, the
longitudinal part having been eliminated through phonon integration.
It can be
written as $R(q)= b N(q) + \mu M(q)$ with:
$$N_{\alpha \beta, \gamma \delta}
= {1 \over {D-1}} P^T_{\alpha \beta} P^T_{\gamma \delta}~~,~~
M_{\alpha \beta, \gamma \delta}= {1 \over 2} ( P^T_{\alpha \gamma}
P^T_{\beta \delta } + P^T_{\alpha \delta} P^T_{\beta \gamma} )
- N_{\alpha \beta, \gamma \delta}
\eqno(3)
$$
where $P^T_{\alpha \beta}=\delta_{\alpha \beta} - q_{\alpha}q_{\beta}/q^2$
is the transverse projector. $\mu$ is the shear
modulus and $b= \mu ( 2 \mu + D \lambda)/(2 \mu + \lambda)$ is proportional
to both shear and bulk moduli.  The convenience of this decomposition
is that $M$ and $N$ are mutually orthogonal projectors
under tensor multiplication
(e.g $M_{\alpha \beta, \gamma \delta}
M_{\gamma \delta, \mu \nu}=
M_{\alpha \beta, \mu \nu}$ etc...).

\bigskip

We set up two coupled integral equations for the propagator of the ${\bf h}$
field and for the
renormalized four point interaction.
We want to evaluate $<h_i(-k) h_j(k)> = \delta_{ij} G(k)$
with $G^{-1}(k)=\kappa_R(k)k^4=\kappa k^4 + \sigma(k)$
where $\sigma(k)$ is the self energy.
The SCSA is defined in diagrammatic form
by the graphs of Fig. 1a and 1b,
where the double solid line denotes the dressed propagator $G(q)$, the dotted
line
the bare interaction $R(q)$ and the wiggly line the "screened" interaction
$\tilde{R}(q)$ dressed by the vacuum polarization bubbles.
We thus obtain two equations, one for $\sigma(k)$ which determines $\eta$, the
other for $R$ which determines $\eta_u$:
$$
\sigma(k)= {2 \over d_c} k_{\alpha}k_{\beta}k_{\gamma}k_{\delta}
\int dq \tilde{R}_{\alpha \beta, \gamma \delta}(q) G(k-q)
\eqno(4a)
$$
$$
\tilde{R}(q) = R(q) - R(q) \Pi(q) \tilde{R}(q)
\eqno(4b)
$$
where
$\Pi_{\alpha \beta, \gamma \delta}(q)
= \int dp p_{\alpha} p_{\beta} p_{\gamma} p_{\delta}
G(p) G(q-p)$ is the vacuum polarization
and tensor multiplication is defined above. Because of the transverse
projectors, only the component
$\Pi(q)_{sym}  $ of $\Pi(q)$ proportional to
the fully symmetric tensor $S_{\alpha \beta, \gamma \delta}=
\delta_{\alpha \beta} \delta_{\gamma \delta} +
\delta_{\alpha \gamma} \delta_{\beta \delta}+
\delta_{\alpha \delta} \delta_{\beta \gamma}$ contributes in (4b).
Defining $\Pi(q)_{sym} = I(q) S$, simple algebra gives
$\tilde{R}(q)=\tilde{\mu}(q) M + \tilde{b}(q) N$ with
renormalized shear and shear-bulk moduli, and the new equations:
$$
\tilde{\mu}(q)= {\mu \over { 1 + 2 I(q) \mu }}
{}~~~\tilde{b}(q)= {b \over {1 + (D+1) I(q) b }}
\eqno(5a)
$$
$$
\sigma(k)={2 \over d_c} \int dq {{\tilde{b}(q) + (D-2) \tilde{\mu}(q) }
\over{D-1}}
(k P^T(q) k)^2 G(k-q)
\eqno(5b)
$$

\bigskip

We now solve these equations in the long-wavelength limit. Substituting
$G(k) \sim \sigma^{-1}(k) \sim Z/ k^{4-\eta}$ in (5a,b), with $Z$ a
non-universal amplitude, we find that
the vacuum polarization integral diverges as:
$$
I(q)\sim Z^2 A(D,\eta) q^{-\eta_u}
\eqno(6)
$$
where $\eta_u=4-D-2 \eta$ is the anomalous exponent of phonons.
Substituting in (5a,b), and defining the amplitude:
$\int dq q^{\eta_u} (k-q)^{-(4-\eta)} (k P^T(q) k)^2
= B(D,\eta) k^{4-\eta}$, one finds (for $\mu,b>0$) that the $Z$ and
$k^{4-\eta}$ factors
cancel and
that $\eta$ is determined self-consistently by the equation for the ${\it
amplitude}$:
$ d_c={D \over {D+1}} {B(D,\eta) \over A(D,\eta)}$, which
after calculation of the integrals defining A,B gives:
$$
d_c={2 \over \eta}
D(D-1) { { \Gamma[1+{1\over 2}\eta] \Gamma[2-\eta] \Gamma[\eta+D]
\Gamma[2-{1\over 2}\eta] }
\over {\Gamma[{1\over 2}D + {1\over 2}\eta] \Gamma[2-\eta-{1\over 2}D]
\Gamma[\eta+{1\over 2}D] \Gamma[{1\over 2}D+2-{1\over 2}\eta]}}
\eqno(7)
$$
For $D=2$ this equation can be simplified, and one finds (Fig. 2):
$$
\eta(D=2,d_c)={ 4 \over {d_c + \sqrt{16 - 2 d_c + d_c^2}}}
\eqno(8)
$$
Thus for physical membranes we obtain: $\eta=0.821$, $\eta_u=0.358$ and:
$$
\zeta=1-{\eta \over 2}={ \sqrt{15} - 1 \over{ \sqrt{15} + 1}} = 0.590..
\eqno(9)
$$
roughly at midvalue of the
present numerical simulations. From (5)
we also obtain $\mathop{{\rm Lim}}\limits_{q\rightarrow 0}
\tilde{\lambda}(q)/\tilde{\mu}(q) ={- 2 \over {D+2}} $
(i.e a negative Poisson ratio).

\bigskip

Expanding the result (7) in $1/d_c$ one obtains:
$$
\eta={8 \over d_c} {D-1 \over D+2} {\Gamma[D] \over {\Gamma[{D \over 2}]^3
\Gamma[2-{D \over 2}]}}+O({1 \over {d_c^2}})
= {2 \over d_c}+O({1 \over {d_c^2}}) ~~~~(for D=2)
\eqno(10)
$$
which coincides with the exact result$^{6,7}$, as expected by construction of
the SCSA.
Similarly, expanding (7) to first order in $\epsilon=4-D$ one finds:
$$
\eta={\epsilon \over {2 + {d_c/12}}}
\eqno(11)
$$
also in agreement with the exact result$^{6,7}$. This is not a general property
of SCSA. Here
it can be traced to the vertex and box diagrams of Fig. 1c being {\it
convergent}.
Indeed, because of the transverse projectors in (2-3) one can always extract
one
power of external momentum from each external $h$ legs, which lowers the degree
of divergence
from naive power counting.
As a result, if one decouples the 4-point vertex $R$ via a mediating field, the
only counterterms needed are for
two-point functions.

\bigskip

We have analyzed the crumpling transition of phantom membranes by the
same method, applied to the isotropic theory of
Ref.18. The exponent $\eta=\eta_{cr}$ at the transition is determined
by$^{16}$:
$$
d=
{ { D(D+1)(D-4+\eta)(D-4+2\eta)(2D-3+2 \eta)\Gamma[{1\over 2}\eta]
\Gamma[2-\eta] \Gamma[\eta+D] \Gamma[2-{1\over 2}\eta] }
\over {2(2-\eta)(5-D-2\eta)(D+\eta-1)\Gamma[{1\over 2}D + {1\over 2}\eta]
\Gamma[2-\eta-{1\over 2}D]
\Gamma[\eta+{1\over 2}D] \Gamma[{1\over 2}D+2-{1\over 2}\eta]}}
$$
$$
\eqno(12)
$$
At the transition the radius of gyration
scales as $R_G\sim L^{\nu}$ with $\nu=(4-D-\eta_{cr})/2$. For $d=3$ and $D=2$
we find $\eta_{cr}=0.535$ and $\nu=0.732$
(Haussdorf dimension $d_H=2.73$). The embedding dimension
$d_u(D)$ above which self-avoidance is ${\it irrelevant}$ for the membrane
${\it at}$ the
crumpling transition is determined by the condition
$d_u=4D/(4-D-\eta_{cr}(d_u))$. Using (12) we find that $d_u(2)=4.98$.

\bigskip

The present method gives interesting predictions for lower critical dimensions.
In the flat phase,
orientational order (i.e in $\nabla {\bf h}$ ) disappears for $D<D_{lc}$,
where $2-\eta(D_{lc},d_c)=D_{lc}$.
{}From (7) this is equivalent to $d_c=D_{lc} (D_{lc}-1)/(2-D_{lc})$. On the
other hand, the
lower critical dimension $D'_{lc}(d)$ for the crumpling transition is defined
by $2-\eta_{cr}(D'_{lc},d)=D'_{lc}$, or equivalently from (12),
$d=D'_{lc}/(2-D'_{lc})$. Since
$d=D/(2-D)$ is clearly equivalent to $d_c=D(D-1)/(2-D)$
we find that the lower critical dimensions of the crumpling transition and of
the flat phase,
as predicted by SCSA, are identical, and given by $D_{lc}(d)=2 d/(1+d)$. Since
they originate
from very different calculations, this indicates that the
SCSA is quite consistent. For codimension 1 manifolds $D_{lc}=\sqrt{2}$
and for fixed embedding space $d=3$, $D_{lc}=3/2$. $D_{lc}$ increases from
$D_{lc}=1$ for $d_c=0$, to
$D_{lc}=2$ when $d_c \rightarrow \infty$ as expected. Note that for $d>3$
self-avoidance cannot modify
the above results, while for $d<3$ it is an open question.

\bigskip

We can compare (8,12) with recent simulations$^{19}$ of $D=2$ membranes
with self-avoidance in higher $d_c$. The membranes are found flat in $d=3,4$
with
$\zeta(d=3)=0.64 \pm 0.04$, $\zeta(d=4)=0.77 \pm 0.04$,
whereas we obtain $0.59$, $0.67$, respectively. The membrane is crumpled
in $d=5$ with $\nu=0.8 \pm0.06$, although $d=5$ seems almost marginal, whereas
we find $\nu=0.8$ {\it at} the crumpling transition where self avoidance is
irrelevant,
although almost marginally so.

\bigskip

Flat membranes with random spontaneous curvature are described by adding the
term $- \int d^Dx {\bf c}(x).\nabla^2 {\bf h}(x)$
in the energy (1), where {\bf c}(x) are Gaussian quenched random
variables$^{10}$.
Within a replica symmetric SCSA, we find a marginally unstable $T=0$ fixed
point,
i.e a long-wavelength solution only if $T \rightarrow 0$ first. Defining the
replica connected and
off-diagonal exponents $\eta$, $\eta'$,
by $\overline{<{\bf h}(-q){\bf h}(q)>_c} \sim q^{-(4-\eta)}$,
$\overline{<{\bf h}(-q){\bf h}(q)>} \sim q^{-(4-\eta')}$
we find$^{16}$ at this fixed point:
$\eta^\prime = \eta$,
$\eta(d_c,D) = \eta_{pure}(4 d_c,D)$.
Thus one can simply replace $d_c$ in the pure result by $4 d_c$!
Again this agrees with the $1/d_c$ and $\epsilon$ expansions$^{10}$.
For physical membranes $D=2$, $d_c=1$, we find from (8) :
$$
\eta= 2/(2+\sqrt{6})=0.449~~~\zeta=0.775
$$
comparing well with the numerical simulation$^{10}$ result $\zeta=0.81 \pm
0.03$.
By analogy with the random field problem$^{20}$, it is quite possible that the
equality $\eta=\eta^\prime$, conjectured in Ref. 10 to all orders, be
corrected when replica-symmetry breaking is included.

\bigskip

In conclusion, we have presented a self-consistent theory of
polymerized membranes which becomes exact in three limits (large $d_c$,
small $\epsilon=4-D$, and $d_c=0$). By construction, it satisfies the exponent
relations $\eta_u=4-D-2\eta$ and$^{16}$ $1/\nu'=D-2+\eta$. These relations are
exact in the true theory because of rotational invariance
$^{6,7}$. It thus predicts
$\nu^{\prime}=1.218$
and $\delta^{\prime}=1.436$ for the buckling transition exponents$^{6}$.
It contradicts the conjecture$^{2}$ $\zeta=1/2$.

\bigskip

We thank D. Nelson, M. Mezard for discussions. PLD acknowledge support from NSF
grant DMS-9100383 and LR from the
Hertz Graduate Fellowship.

\newpage

\bigskip

\begin{center}
\begin{large}
{\bf References}
\end{large}
\end{center}
\bigskip

*Also LPTENS, Ecole Normale Superieure,
24 rue Lhomond, Paris 75231 Cedex 05, Laboratoire Propre du CNRS.

\bigskip

\begin{enumerate}

\item
See, e.g, Statistical Mechanics of Membranes and Interfaces, edited by D.R.
Nelson,
T. Piran, S. Weinberg ( World Scientific, Singapore 1988 ), and
S. Leibler in Proceedings of the Cargese school on biologically inspired
physics, (1990), to be published.
\item
R. Lipowsky and M. Girardet, Phys. Rev. Lett. ${\bf 65}$ 2893 (1990).
\item
M. Mutz, D. Bensimon, M.J. Brienne, Phys. Rev. Lett. ${\bf 67}$ 923 (1991).
\item
X. Wen et al. Nature {\bf 355}, 426 (1992)
\item
D. R. Nelson and L. Peliti, J. Phys. (Paris) ${\bf 48}$, 1085 (1987).
\item
F. David and E. Guitter, Europhys. Lett. ${\bf 5 }$, 709 (1988).
E. Guitter, F. David, S. Leibler and L. Peliti, J. Phys. France ${\bf 50}$ 1787
(1989).
\item
J.A. Aronovitz and T.C. Lubensky, Phys. Rev. Lett. ${\bf 60}$, 2634 (1988),
J.A. Aronovitz, L. Golubovic, T.C. Lubensky, J. Phys. France {\bf 50} 609
(1989).
\item
F.F. Abraham, D.R. Nelson, J. Phys. France ${\bf 51}$ 2653 (1990).
F.F. Abraham, W.E. Rudge and M. Plishke, Phys. Rev. Lett. ${\bf 62}$, 1757
(1989).
\item
D.R. Nelson, L. Radzihovsky, Europhys. Lett. ${\bf 16}$, 79 (1991),
L. Radzihovsky, P. Le Doussal, J. Phys. I France ${\bf 2}$, 599 (1992).
\item
D.C. Morse, T.C. Lubensky, G.S. Grest, Phys. Rev A ${\bf 45}$ R2151 (1992).
Morse Lubensky, Preprint 1992.
\item
F. Abrahams, Phys. Rev. Lett. ${\bf 67}$, 1669 (1991).
\item
S. Leibler and A. Maggs, Phys. Rev. Lett. ${\bf 63}$, 406 (1989).
\item
G. Gompper and D.M. Kroll, J. Phys. I France ${\bf 2}$, 663 (1992).
(1992).
\item
J. Toner, Phys. Rev. Lett. {\bf 64} 1741 (1990).
\item
A.J. Bray, Phys. Rev. Lett. ${\bf 32}$, 1413 (1974).
\item
P. Le Doussal, L. Radzihovsky, to be published.
\item
E.D. Siggia, B.I. Halperin and P.C. Hohenberg, Phys. Rev. B {\bf 13} 2110
(1976).
\item
M. Paczuski, M. Kardar and D.R. Nelson, Phys. Rev. Lett. ${\bf 60}$, 2638
(1988).
\item
G. Grest, J. Phys. I France {\bf 1},1695 (1991).
\item
M. Mezard, A.P. Young, Preprint LPTENS 92/2

\end{enumerate}
\newpage

\begin{center}
{\bf FIGURE CAPTIONS} \\
\end{center}
\bigskip

\begin{description}

 \item [Figure 1] :
graphical representation of the SCSA: (a) self energy, (b) interaction. (c) UV
finite vertex and box diagrams.

\item [Figure 2] :
$\zeta$ as a function of $d$ for two-dimensional membranes $D=2$. Solid curve:
SCSA result
(8). Dashed dotted curve: $O(\epsilon)$ result, setting
$\epsilon=2$. Dashed curve: corresponds to $\eta=2/d$ chosen (somewhat
arbitrarily) in Ref. 6
as a possible interpolation to finite $d$ (asymptotic to the solid curve for $d
\rightarrow
\infty$).

\end{description}

\end{document}